# NONLINEAR SELF-ACTION ACROSS TEMPORAL REGIMES IN RESONANT DIELECTRIC METASURFACES

**ALFONSO NARDI,[1] SONIA FREDDI,[2] MICHAEL SCALORA,[3] AGOSTINO DI FRANCESCANTONIO,[1] JOHANN OSMOND,[4] SOFIA MARTINS,[4] ATTILIO ZILLI,[1] MARCO FINAZZI,[1] MICHELE CELEBRANO,[1]* MONICA BOLLANI,[2]* MARIA ANTONIETTA VINCENTI[5]**

*[1]Department of Physics, Politecnico di Milano, Piazza Leonardo 32, 20133 Milano, Italy*
*[2] Institute of Photonic and Nanotechnology (IFN) - Consiglio Nazionale delle Ricerche (CNR), Piazza Leonardo 32, 20133 Milano, Italy*
*[3]Institute of Photonics and Quantum Sciences, Heriot-Watt University, SUPA, Edinburgh EH144AS, United Kingdom*
*[4] ICFO - Institut de Ciències Fotòniques, the Barcelona Institute of Science and Technology, Castelldefels, Barcelona 08860, Spain*
*[5]Department of Information Engineering, Università degli Studi di Brescia, Via Branze 38, 25123 Brescia, Italy*
**Correspondence: Monica Bollani (monica.bollani@cnr.it) or Michele Celebrano (michele.celebrano@polimi.it)*

## ABSTRACT

Quasi-bound states in the continuum (qBICs) enable exceptional field confinement, strongly reducing the pump intensity threshold for nonlinear light–matter interaction in dielectric metasurfaces. As a result, nonlinear self-action effects, often elusive in bulk nonlinear media, emerge at moderate excitation intensities. Here, nonlinear self-action in resonantly enhanced third-harmonic (TH) generation from a dielectric metasurface supporting a qBIC resonance is investigated across distinct temporal excitation regimes. These regimes establish different coupling conditions between the excitation and the resonant mode, causing the same nonlinear self-action to emerge through complementary intensity-dependent signatures. Under spectrally narrow picosecond excitation, resonance-enhanced TH generation shows pronounced deviations from cubic scaling at high intensities. In contrast, broadband femtosecond excitation transiently drives the resonance, encoding the nonlinear response in the spectral reshaping and broadening of the TH signal. Simulations reproduce both regimes: continuous-wave modeling captures picosecond power scaling and the role of higher-order nonlinear susceptibilities, while time-domain simulations resolve femtosecond dynamics. These results clarify how pulse duration, bandwidth, and resonant coupling determine the observable signatures of nonlinear self-action in resonant dielectric metasurfaces, linking field confinement to conversion efficiency, power-law scaling, and ultrafast spectral dynamics.

## INTRODUCTION

Nonlinear light–matter interactions provide a rich and versatile toolkit for applications such as frequency conversion[1-4], ultrafast optical control[5-7] and all-optical switching[8-12]. However, the intrinsically weak nonlinear response of most optical materials typically necessitates either long interaction lengths or high optical intensities, constraining device miniaturization and overall efficiency. Optical nanostructures offer a powerful route to overcome these limitations by reducing modal volumes and increasing field localization, thereby strengthening nonlinear interactions at the nanoscale level[13]. In particular, dielectric nanostructures supporting resonant modes have emerged as especially promising platforms, combining strong optical confinement with low absorption losses and high damage thresholds[14].

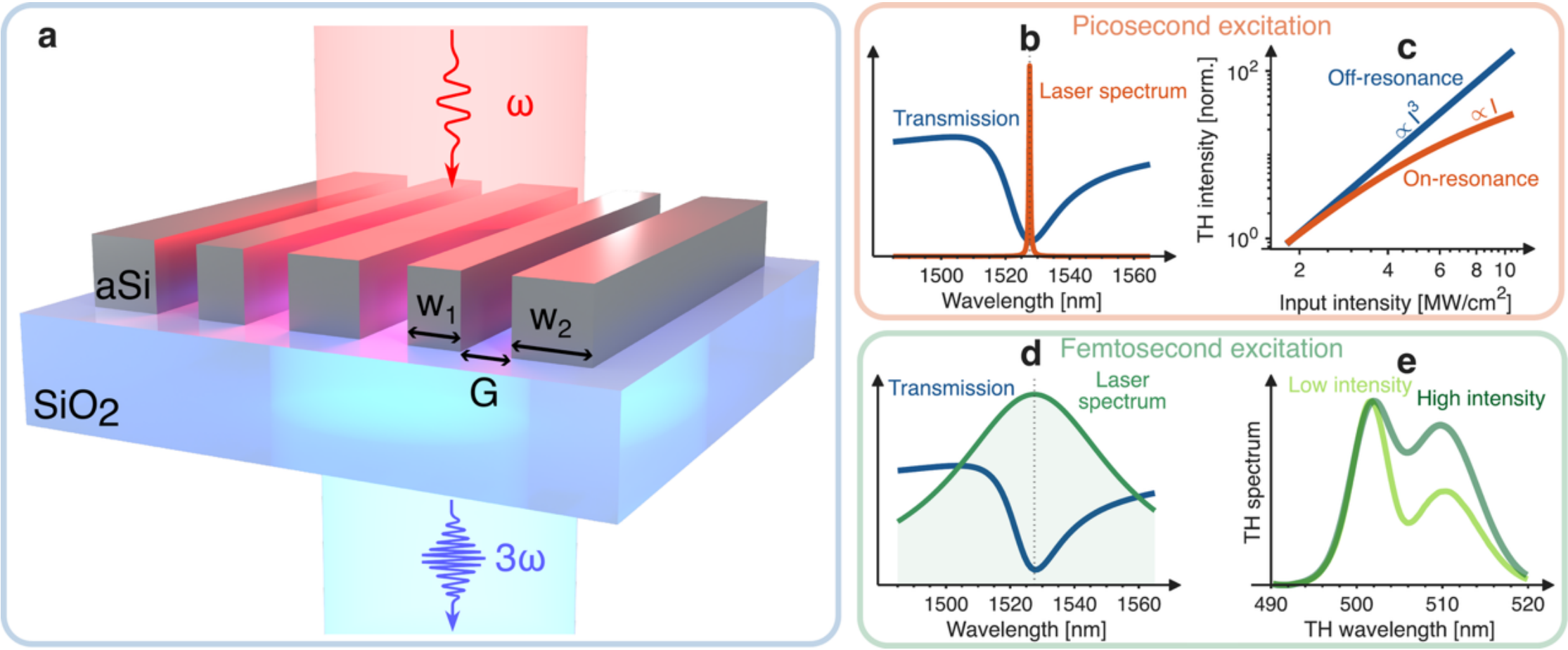

**Figure 1 – Conceptual overview of third-harmonic generation in qBIC metasurfaces under picosecond and femtosecond excitation**. (a) Schematic of the asymmetric dielectric metasurface supporting a qBIC. The unit cell consists of two nanobars of different widths ($w_1$, $w_2$) separated by a gap (G) on a $SiO_2$ substrate. Excitation at the fundamental frequency $\omega$ leads to resonantly enhanced third-harmonic (TH) emission at $3\omega$ due to strong field confinement within the nanobars. (b,c) Picosecond excitation regime: the narrow laser spectrum is much smaller than the qBIC linewidth, resulting in selective excitation of the resonance. The TH intensity follows cubic scaling off resonance, while near resonance nonlinear self-action leads to sub-cubic behavior. (d,e) Femtosecond excitation regime: the broadband pump spans the resonance, and only a portion of the spectrum efficiently couples to the qBIC. The resulting TH emission exhibits spectral filtering by the resonance. At high intensity, nonlinear self-action induces spectral reshaping and broadening of the TH signal.

Metasurfaces supporting bound states in the continuum (BICs) were initially explored due to their theoretically unbounded quality factors[15-19]. Over time, attention shifted to more practical implementations in which symmetry breaking or radiative coupling converts ideal BICs into quasi-bound states in the continuum (qBICs). These resonances can retain high quality factors while remaining accessible to free-space excitation[20]. Their extreme electric-field enhancement and prolonged photon lifetimes naturally position them as ideal candidates for boosting nonlinear optical processes[14,21-23].

Recent studies have demonstrated efficient harmonic generation[24-26], enhanced Kerr nonlinearities[27,28], and low-threshold nonlinear effects in qBIC-based dielectric metasurfaces[29], emphasizing their potential for compact and highly efficient nonlinear photonic devices. However, while resonant field confinement offers an exceptional route to enhancing the efficiency of nonlinear processes, narrow resonances can also become highly sensitive to nonlinear self-action effects and to the temporal and spectral characteristics of the incident excitation. Optical nonlinearities, most notably the Kerr effect, induce intensity-dependent refractive-index changes that might modify the spectral features of a resonant system. This mechanism can lead to nonlinear resonance shifts, saturation of harmonic generation processes, deviations from simple power-law scaling, and complex spatio-temporal dynamics[30]. In resonant nanostructures, even small refractive-index variations can strongly perturb the resonant conditions, making self-action effects not merely secondary corrections but potentially dominant contributors to the overall nonlinear behavior.

Viewed from this perspective, nonlinear self-action in resonant metasurfaces can also be understood as a form of intrinsic temporal modulation: the optical response of the structure evolves during the excitation itself as the resonant field modifies the refractive index. This connection places qBIC nonlinear dynamics within the broader context of time-varying photonic media, a field that has recently attracted growing interest for realizing nonreciprocity, frequency translation, parametric amplification, and ultrafast optical control[31-35]. Importantly, such time-dependent behavior does not require a conceptual departure from established electromagnetic theory. The relevant dynamical ingredients are already contained, often in their most general form, within the Maxwell–Drude–Lorentz and Maxwell–Bloch frameworks, which provide a complete and self-consistent description of the temporal evolution of polarization, dispersion, and energy exchange between fields and matter.

Despite growing experimental and theoretical interest, nonlinear self-action in qBIC metasurfaces has thus far been examined primarily within specific excitation regimes, often without fully considering how the coupling regime evolves in the presence of an intensity-dependent resonant response. Previous studies have demonstrated nonlinear self-action, resonance reshaping, and deviations from conventional power-law scaling in resonant metasurfaces[29, 30]. However, a direct comparison of distinct temporal excitation regimes interacting with the same resonance remains largely unexplored. In particular, pulse duration and the corresponding excitation bandwidth determine whether the resonant mode is driven in a quasi-stationary or transient manner. Narrowband excitation can efficiently drive a resonance over its characteristic lifetime, leading to a quasi-stationary interaction, whereas broadband excitation interacts with it transiently and in a spectrally selective manner. How these distinct coupling scenarios, established by different temporal excitation regimes, determine the observable signatures of nonlinear self-action remains an open question.

Here, we address this question by systematically investigating nonlinear self-action across two distinct temporal excitation regimes in a dielectric metasurface supporting a qBIC resonance, using TH generation as our nonlinear probe (Figure 1a). By comparing narrowband picosecond and broadband femtosecond excitation of the same resonant platform, we show that nonlinear self-action manifests through complementary observables in the two regimes. Under picosecond excitation (Figure 1b), the narrowband pulse efficiently couples to the resonance, leading to strong enhancement of TH emission. In this regime, we observe clear deviations from the cubic power scaling expected for a third-order nonlinear process, ultimately leading to saturation at high intensities, as schematically illustrated in Figure 1c. These observations directly reveal how resonance-mediated self-action modifies and limits the nonlinear conversion efficiency. In contrast, when the metasurface is driven by broadband femtosecond pulses (Figure 1d), its interaction with the qBIC occurs in a transient regime. In this case, the nonlinear response manifests most clearly as a power-dependent reshaping of the emitted TH spectrum, rather than solely as a change in conversion efficiency (as sketched in Figure 1e). This behavior reflects the dynamic interplay between the excitation bandwidth and the ultrafast evolution of the resonant mode, giving rise to distinct spectral signatures of nonlinear self-action.

To model the nonlinear response in the picosecond regime, we employ spectrally resolved continuous-wave (CW) simulations, justified by the narrow excitation linewidth, which fully resolves the qBIC resonance and makes the interaction effectively quasi-stationary. This approach captures the observed TH enhancement, resonance reshaping, and deviations from cubic scaling, while also providing direct physical insight into the role of higher-order nonlinear susceptibilities. In contrast, under femtosecond excitation the laser linewidth exceeds the qBIC linewidth, so that the interaction with the resonance is intrinsically transient. In this case, the nonlinear response is encoded in the ultrafast spectral evolution of the emitted TH signal and requires full time-domain simulations based on a hydrodynamic Maxwell-Lorentz framework. Together, the two numerical approaches show how the same field-confinement-driven mechanism governs both efficiency and power-law scaling under narrowband excitation, while simultaneously exerting a decisive influence on spectral dynamics under broadband excitation.

A central message emerging from this work is that ordinary materials, when driven into regimes of strong field confinement and ultrafast nonlinear response, naturally operate as time-varying systems. The spectral reshaping of the resonance, transient coupling dynamics, and power-dependent spectral signatures we observe do not require engineered temporal modulation or exotic material platforms. They arise directly from the intrinsic physics of resonant dielectric nanostructures.

This perspective underscores two important points. First, the time-dependent behavior is an inherent consequence of the nonlinear, strongly resonant regime, not of an externally imposed modulation scheme. Second, the search for new spatio-temporal phenomena should not overlook the rich, built-in dynamics conventional materials already exhibit when pushed into ultrafast, high-field conditions.

By placing qBIC-mediated self-action within this broader conceptual framework, our work highlights that time-dependence is not an add-on to such systems: it is embedded in their fundamental operation.

Clarifying how distinct temporal excitation regimes, through their different bandwidths and coupling conditions, shape nonlinear self-action advances the foundational understanding of nonlinear dynamics in resonant metasurfaces and provides concrete guidance for the design and experimental exploration of ultrafast and nonlinear nanophotonics.

## RESULTS

### Metasurface design and linear optical response

We investigate a one-dimensional periodic array of amorphous silicon nanobars patterned on a fused-silica substrate, with a deliberately introduced in-plane asymmetry within each unit cell[36,37]. As shown in Figure 2a, each period contains two nanobars of identical height (140 nm) but different widths ($w_1$ = 147 nm and $w_2$ = 190 nm), separated by a 255 nm gap G, with a lattice period of 770 nm. The fabricated structure, shown in the SEM image (Figure 2b), confirms the intended periodic geometry and the controlled in-plane asymmetry of the unit cell (fabrication details are provided in the Materials and methods section). This controlled symmetry breaking enables the excitation of a delocalized qBIC when the structure is illuminated with transverse-electric (TE) polarized light, i.e., with the electric field oriented parallel to the nanobars[38]. The corresponding resonant mode exhibits a spatially structured near-field distribution, as evidenced by the simulated $E_z/E_0$ profile obtained from COMSOL Multiphysics (Figure 2c), where $E_z$ denotes the electric field component parallel to the nanobars, showing opposite-phase field components localized on the two asymmetric nanobars. This field distribution is accompanied by a pronounced electromagnetic field enhancement, arising from the suppressed radiative coupling that gives rise to the qBIC resonance.

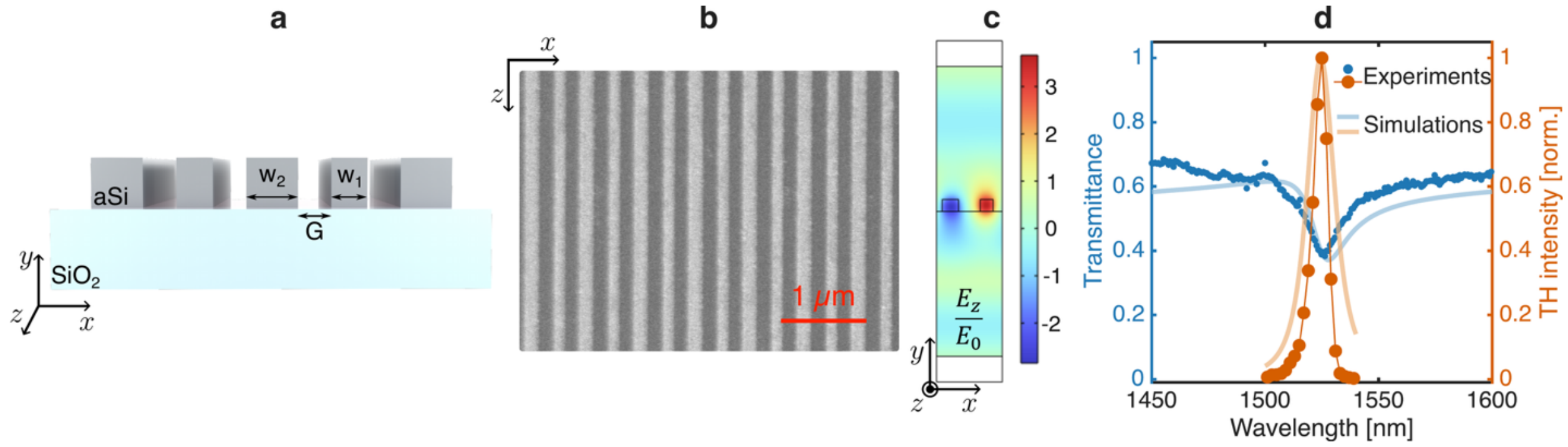

**Figure 2 – Spectral resonance and nonlinear response of the designed metasurface.** (a) Schematic of the asymmetric dielectric nanobar unit cell designed to support a qBIC ($w_1$ = 147 nm, $w_2$ = 190 nm, G = 255 nm, unit cell period of 770 nm). The in-plane symmetry breaking enables coupling to the otherwise dark mode under TE-polarized excitation (i.e., electric field oriented parallel to the ridges). (b) Scanning electron microscope (SEM) image of the fabricated metasurface, confirming the periodic nanobar geometry and the designed structural asymmetry. (c) Simulated TE electric field distribution ($E_z/E_0$), with $E_z$ parallel to the nanobars, at the qBIC resonance, showing a spatially structured near-field profile with opposite-phase components localized on the two nanobars. (d) Linear transmission spectrum (blue) and normalized TH intensity (orange) under TE-polarized excitation. Experimental data are shown as markers, and numerical simulations as solid lines.

We characterized the linear optical response of the metasurface by measuring its transmission spectrum under broadband illumination from a supercontinuum source, with the transmitted signal analyzed using an infrared spectrometer (see the Materials and methods section and Sec. S1 of the Supplementary Information for details on the linear-transmission measurement and experimental setup). Under TE-polarized excitation, the transmission spectrum exhibits a pronounced asymmetric Fano resonance, which is the spectral signature of the qBIC mode (Figure 2d). In contrast, when the metasurface is illuminated with transverse-magnetic (TM) polarized light, orthogonal to the nanobars, the transmission remains nearly flat over the same spectral range, confirming the polarization-selective nature of the qBIC excitation (see Sec. S2 of the Supplementary Information). The experimentally measured spectral response of the structure (Figure 2d) is in good qualitative agreement with numerical simulations. The remaining quantitative deviations are mainly ascribed to fabrication-induced variations in the critical dimensions, particularly few-nanometer deviations of the inter-bar gap with respect to the nominal design, which are compatible with the tolerance of 50-kV electron-beam lithography. Slightly oblique sidewalls, surface roughness, and other incidental fabrication-related effects may further contribute to the observed discrepancies. The resonant qBIC response leads to strong electromagnetic field confinement within the nanobars, providing the resonant enhancement necessary to efficiently drive the TH generation explored in the following sections, as evidenced in Figure 2d (orange curve), where the normalized TH signal measured under picosecond excitation exhibits a pronounced peak spectrally aligned with the qBIC resonance.

Having established the qBIC resonance and its spectral overlap with enhanced TH generation, we next use TH emission to probe the nonlinear response under two excitation regimes. Picosecond pulses provide narrowband excitation that resolves the qBIC linewidth, enabling wavelength-dependent measurements of conversion efficiency and power scaling. Femtosecond pulses, by contrast, provide broadband excitation that overlaps the resonance transiently, so that the nonlinear response is accessed through the spectral structure of the emitted TH signal.

### Picosecond excitation

We first investigate TH generation from the metasurface under picosecond excitation, for which the spectral bandwidth of the pump is much narrower than the linewidth of the qBIC resonance. The experiments are performed using a tunable picosecond laser (≈7 ps pulse duration, 0.5 nm linewidth, repetition rate of 50 MHz), enabling precise control of the excitation wavelength across the resonance. By sweeping the fundamental wavelength in the range 1505–1537 nm and detecting the generated TH signal with a single-photon avalanche diode, we reconstruct the normalized TH intensity as a function of excitation wavelength, which exhibits a pronounced maximum at the qBIC resonance wavelength (see the Materials and methods section and Sec. S3 of the Supplementary Information for details on the picosecond nonlinear measurements and power-stability characterization).

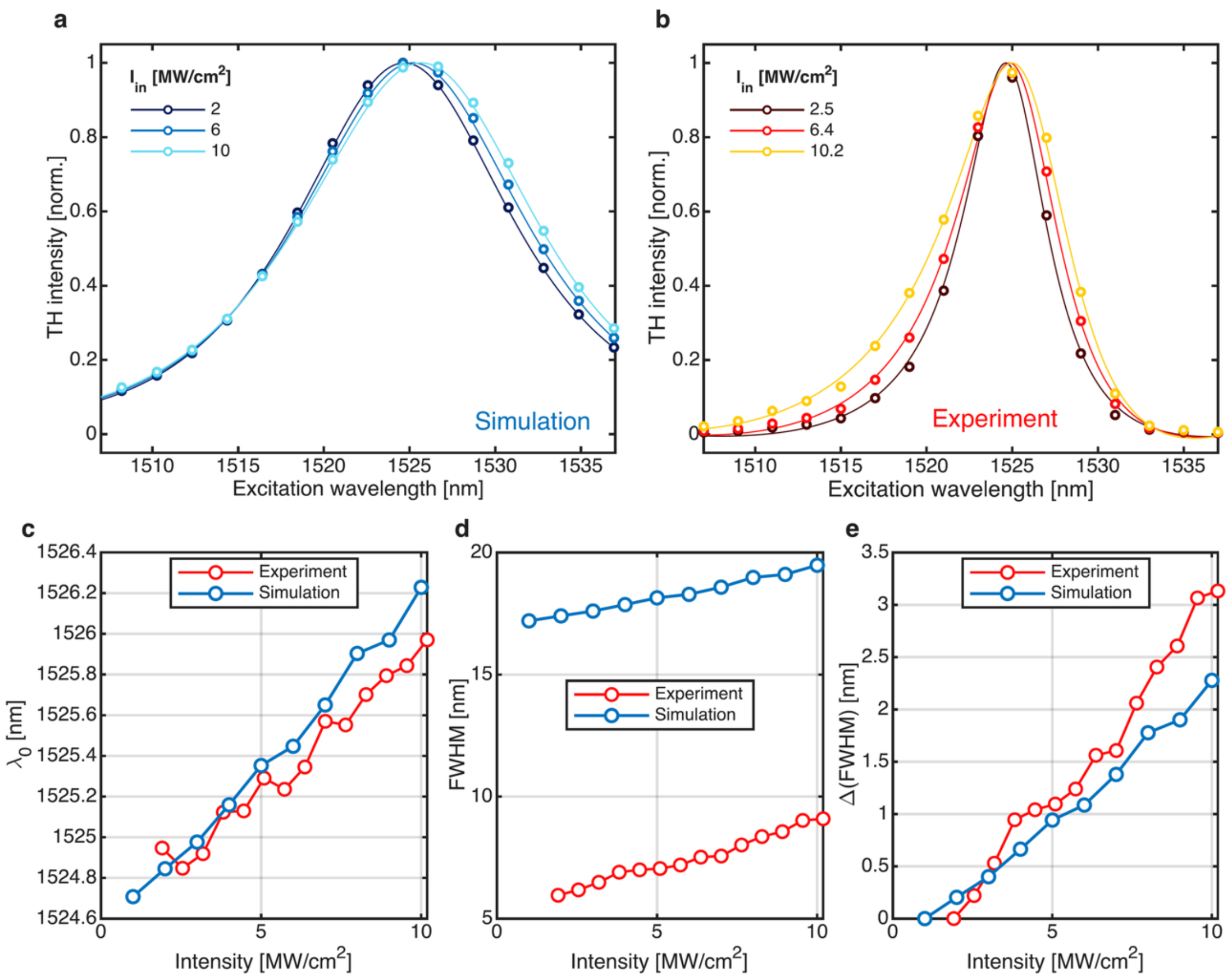

**Figure 3 – Third-harmonic generation under picosecond excitation.** (a,b) Simulated (a) and experimentally measured (b) normalized TH intensity as a function of excitation wavelength for increasing input peak intensities. Markers represent simulation and experimental data, respectively, while solid lines correspond to Fano resonance fits. The normalization is performed with respect to the maximum of each corresponding fit. A progressive spectral redshift and broadening are observed with increasing input power. (c–e) Parameters extracted from the Fano resonance fits as a function of input intensity: (c) central wavelength $\lambda_0$, (d) full width at half maximum (FWHM), and (e) spectral broadening $\Delta(FWHM)$, defined as the increase in FWHM relative to its low-intensity value.

Figure 3a shows the simulated normalized TH intensity as a function of excitation wavelength for increasing input peak intensities (2, 6, and 10 MW/cm$^2$). With increasing intensity, the resonance undergoes a progressive redshift and spectral broadening, reflecting the intensity-dependent modification of the response under resonant excitation.

In the picosecond regime, the excitation bandwidth is significantly narrower than the qBIC linewidth, allowing the resonance to be spectrally resolved and efficiently driven over its characteristic lifetime. Under these conditions, the interaction is effectively quasi-stationary, and the nonlinear response can be accurately described using spectrally resolved CW simulations performed with COMSOL Multiphysics. This approach captures the experimentally observed resonance reshaping and TH power scaling, while also providing a transparent interpretation of the role of higher-order nonlinear susceptibilities.

The simulated data (markers in Figure 3a) are well fitted by a Fano resonance (solid lines), from which we extract the evolution of the resonance parameters with increasing intensity.

Figure 3b shows the transmitted TH spectra for three incident peak power densities (2.5, 6.4, and 10.2 MW/cm$^2$). Motivated by the simulations, the experimental data (markers) are fitted using a Fano lineshape (solid lines). The normalization is performed with respect to the maximum of the fitted resonance, enabling a consistent evaluation of spectral shifts. This procedure avoids artifacts arising from the finite wavelength sampling of the measurements, which would otherwise pin the observed maximum of the TH intensity to a fixed wavelength (≈1525 nm), as shifts smaller than the discretization step (2 nm) cannot be resolved.

Figures 3c–e show the parameters extracted from the Fano fits as a function of input peak intensity, for both experiment and simulations. Figure 3c reports the resonance wavelength, which exhibits a systematic redshift with increasing intensity. Figure 3d shows the full width at half maximum (FWHM), while Figure 3e reports the corresponding spectral broadening, defined as the increase in FWHM relative to its low-intensity value. The redshift and broadening increase approximately linearly over the explored range, and the agreement between experiment and simulation is excellent. To identify the dominant mechanism underlying the spectral shift, we compared simulations performed with and without the two-photon absorption contribution. The nearly identical resonance shifts obtained in the two cases indicate that, within the investigated intensity range, the observed redshift is predominantly driven by the Kerr-induced refractive-index change, whereas nonlinear absorption has only a minor influence on the resonance position (see Supplementary Section S9). The slight discrepancy in the absolute FWHM values (Figure 3d) is attributed to uncertainties in the material parameters, such as the refractive index and third-order susceptibility.

We obtain additional insight into the spectral broadening by experimentally analyzing the intensity dependence of the TH signal as a function of input fundamental intensity (Figure 4). The power-dependent TH intensity for different excitation wavelengths is shown in Figures 4a and 4b for simulations and experiments, respectively. When the pump is detuned from the resonance, the TH intensity follows the expected cubic dependence on excitation power, characteristic of a third-order nonlinear process (Figures 4a,b, black curves). In contrast, as the excitation wavelength approaches the maximum of the resonance-enhanced response, pronounced deviations from cubic scaling emerge. In this regime, the power dependence becomes progressively sub-cubic and can approach nearly linear behavior at the highest intensities investigated, consistent with saturation associated with the nonlinear modification of the qBIC response.

This transition is more clearly visualized in the log–log representation of the experimental data (Figure 4c), where the change in slope at high intensities highlights the departure from cubic scaling. This behavior is quantified in Figure 4d, where we report the fitted power-law exponent $\alpha$ extracted in the high-intensity regime. The individual power-dependent TH intensity curves and the corresponding fits for each excitation wavelength are provided in Sec. S4 of the Supplementary Information. The corresponding picosecond TH efficiency curves as a function of input intensity are provided in Sec. S5 of the Supplementary Information. While all datasets exhibit cubic scaling at low intensities, at high intensities the exponent decreases markedly near resonance and gradually returns toward the cubic value upon detuning, consistent with a self-action–driven modification of the resonant response.

This behavior directly accounts for the spectral broadening observed in Figure 3. Specifically, the on-resonance wavelength does not scale with power as rapidly as the off-resonant wavelengths. Upon normalization of the TH intensity, this reduced scaling effectively enhances the relative contribution of off-resonant spectral components, which retain cubic power dependence, compared to the near-resonant wavelengths, which exhibit sub-cubic scaling, leading to an effective broadening of the normalized TH intensity profile.

The deviation from cubic scaling arises from the increasing contribution of higher-order nonlinear susceptibilities under resonant field enhancement. At sufficiently large local field intensities, the nonlinear response cannot be described by the third-order susceptibility alone, and fifth- and seventh-order contributions must also be included. In the model, the nonlinear polarization is therefore expanded up to $\chi^{(7)}$, with $\chi^{(3)}$, $\chi^{(5)}$, and $\chi^{(7)}$ contributing to the nonlinear polarization at both the fundamental and third-harmonic frequencies. Because these susceptibilities are complex, their combined action modifies both the amplitude and phase of the nonlinear polarization and introduces intensity-dependent dispersive and absorptive contributions, progressively altering the TH power scaling and leading to saturation at high excitation intensities. Importantly, only the fields at $\omega$ and $3\omega$ are propagated in the numerical model; higher-frequency components such as $5\omega$ and $7\omega$ are not included. Further details of the nonlinear model are provided in the Materials and methods section, and in Sec. S8 of the Supplementary Information.

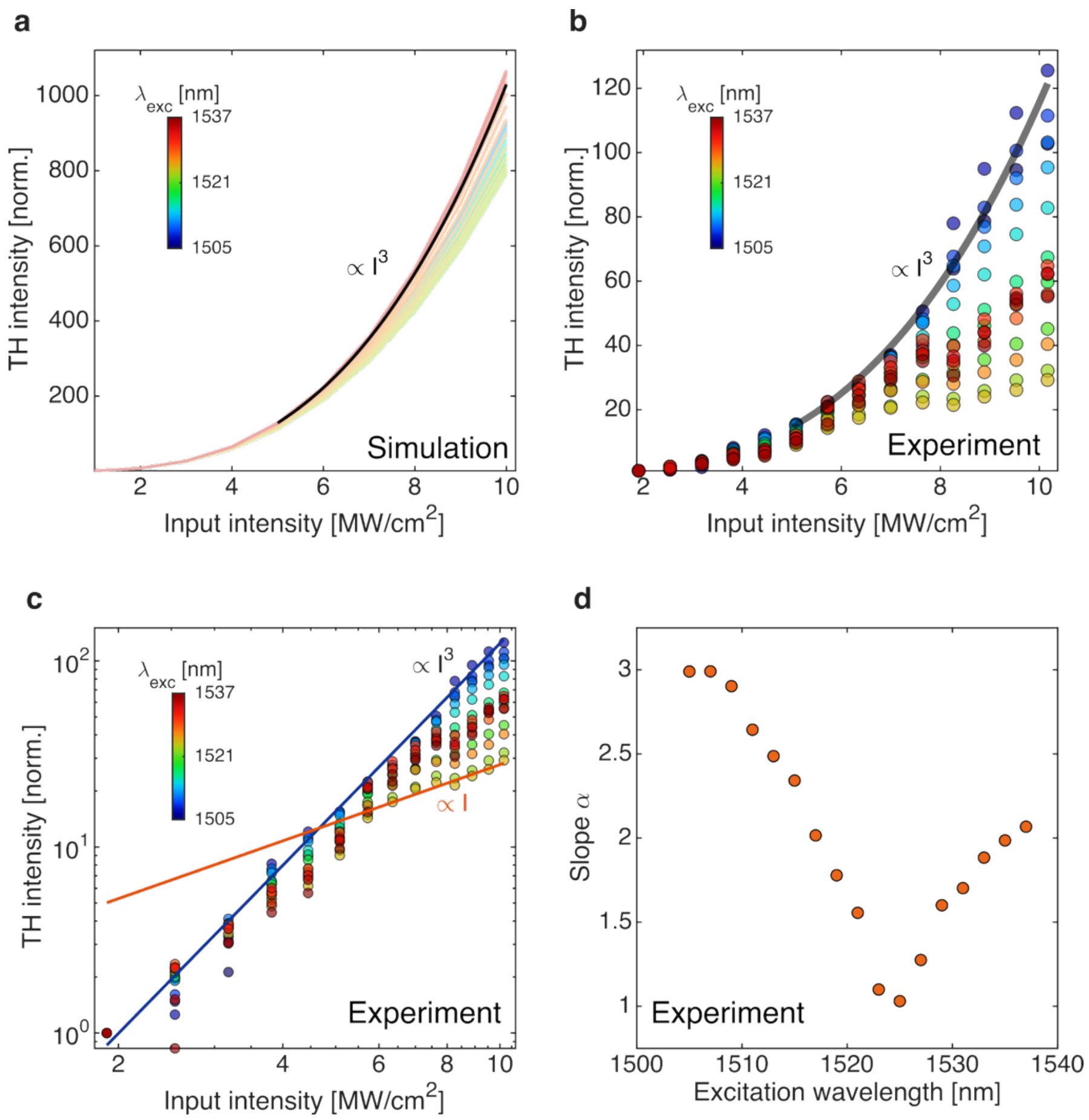

**Figure 4 – Intensity-dependent third-harmonic generation under picosecond excitation**. (a,b) Simulated (a) and experimentally measured (b) TH intensity as a function of input intensity for different excitation wavelengths across the resonance. For each wavelength, the TH intensity is normalized to its value at the lowest measured input intensity. The TH intensity follows a cubic dependence ($\propto I^3$) far from resonance, while the scaling progressively deviates from cubic behavior as the excitation wavelength approaches the resonance maximum. (c) Log–log representation of the experimental data with power-law fits, highlighting the deviation from cubic scaling at high intensities and near-resonant excitation, where the slope approaches nearly linear behavior ($\alpha \approx 1$). (d) Extracted power-law exponent $\alpha$ as a function of excitation wavelength. The exponent decreases as the excitation wavelength approaches the resonance peak, and increases again as the system is detuned from the resonance.

These observations demonstrate that, under spectrally narrow picosecond excitation, where the pulse bandwidth is narrow enough to fully resolve the qBIC resonance and effectively behaves as a quasi-CW drive, the intense field confinement provided by the qBIC simultaneously enhances the TH conversion efficiency and induces nonlinear self-action effects that modify the power scaling of the emitted signal.

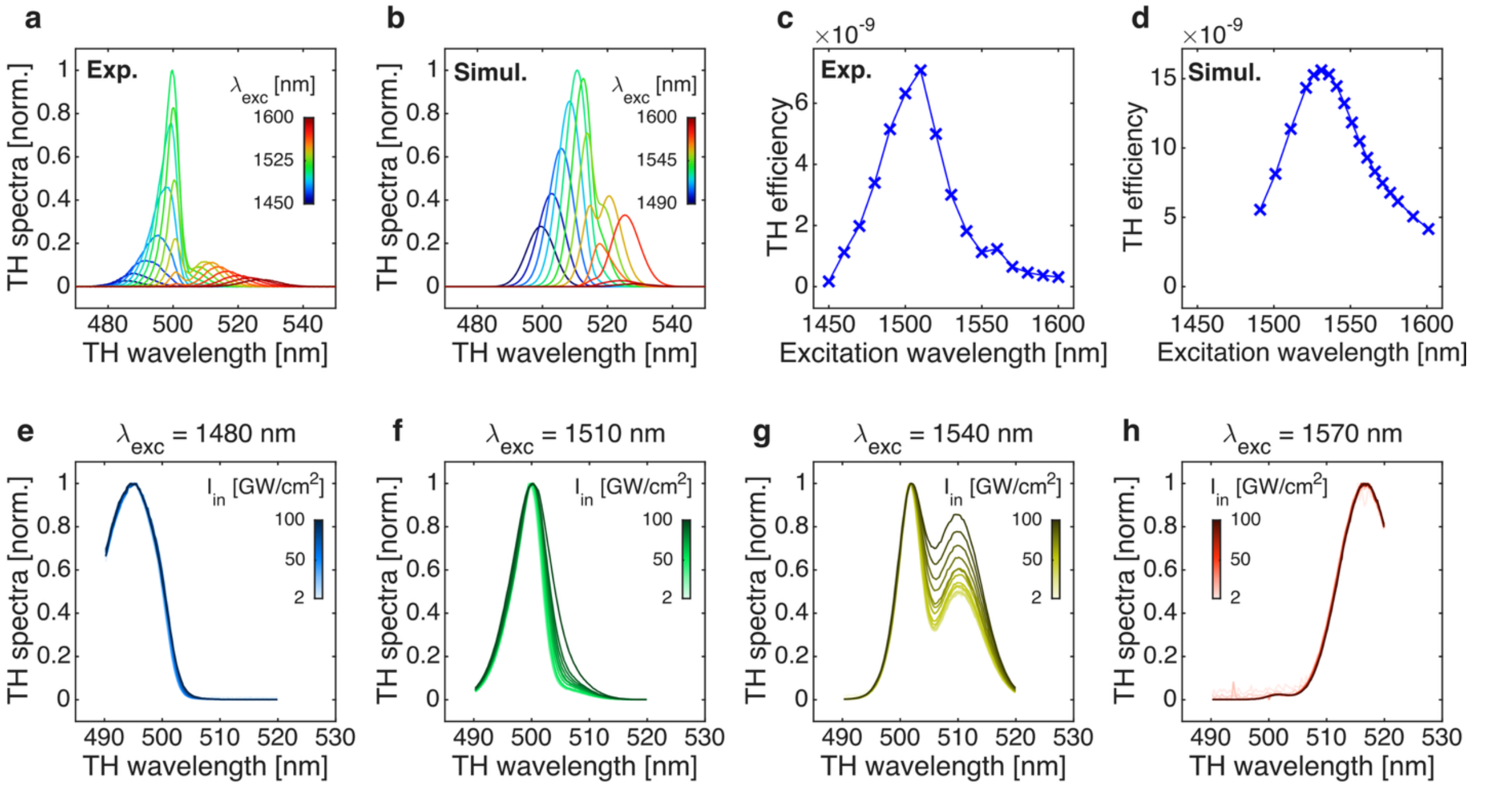

**Figure 5 – Third-harmonic generation under femtosecond excitation.** (a, b) Normalized TH spectra measured (a) and simulated (b) for a series of fundamental wavelengths. The color scale indicates the excitation wavelength of the femtosecond pump. For each wavelength, a femtosecond pulse centered at the corresponding fundamental frequency generates TH emission, which is measured with a spectrometer. (c,d) TH conversion efficiency as a function of excitation wavelength, measured (c) and simulated (d). (e-h) Intensity-dependent TH spectra for selected excitation wavelengths: (e) 1480 nm, (f) 1510 nm, (g) 1540 nm, and (h) 1570 nm. When the pump wavelength lies far from the resonance (e, h), increasing the input power produces negligible spectral distortion. In contrast, near resonance (f, g), the TH spectra broaden significantly with increasing intensity, due to self-action effects.

## Femtosecond excitation

Next, we investigate TH generation under femtosecond excitation. Under picosecond excitation, the narrow pump linewidth allows the qBIC resonance to be resolved by scanning the excitation wavelength, but the emitted TH spectrum itself remains spectrally narrow and essentially dictated by the pump linewidth. Therefore, resonance-induced spectral reshaping of the nonlinear emission cannot be directly accessed within a single spectrum. Under femtosecond excitation, instead, the generated TH signal is collected and spectrally resolved using a spectrometer, providing direct access to the spectral characteristics of the nonlinear emission. The experiments are performed using a tunable femtosecond laser delivering pulses of approximately 60 fs and a spectral bandwidth of about 60 nm (details on the femtosecond nonlinear measurements are provided in the Materials and methods section, while the corresponding experimental setup is shown in Sec. S6 of the Supplementary Information).

In this regime, the spectral bandwidth of the excitation becomes larger than the qBIC linewidth. As a result, multiple spectral components of the femtosecond pulse overlap with the resonance simultaneously and couple to it with different efficiencies, giving rise to a transient resonant response. Consequently, the nonlinear dynamics are encoded in the spectral structure of the emitted TH signal. Capturing these spectral dynamics is addressed using full time-domain simulations based on the hydrodynamic Maxwell-Lorentz model, as described in the Materials and methods section. In contrast, a CW treatment would neglect the finite pulse duration and the associated transient coupling dynamics and therefore cannot reproduce the temporal evolution that shapes the broadband TH spectrum.

Figures 5a and 5b present the normalized TH spectra measured (Figure 5a) and simulated (Figure 5b) as a function of the central wavelength of the femtosecond pump, tuned from 1450 nm to 1600 nm. When the excitation wavelength approaches the qBIC resonance, the TH spectra exhibit pronounced spectral modulation, with distinct features that arise from the narrow resonant response. In contrast, for pump wavelengths far detuned from resonance, the TH spectra remain smooth and closely follow the spectral profile of the incident pulse. The experimental observations (Figure 5a) are accurately reproduced by the simulations (Figure 5b).

The resonant enhancement also results in a substantial increase in conversion efficiency, as evidenced in Figures 5c and 5d for experiment and simulation, respectively. The TH efficiency as a function of excitation wavelength is significantly broader than in the picosecond case, reflecting the convolution of the narrow qBIC resonance with the broad spectral bandwidth of the femtosecond pump. This observation confirms that only a limited portion of the pump spectrum effectively couples to the qBIC mode.

To investigate self-action effects, we increase the excitation intensity from 1.6 to 95 GW/cm$^2$ while keeping the pump wavelength fixed. Due to the reduced spectral overlap with the resonance, the onset of spectral modulation occurs at intensities that are orders of magnitude higher than in the picosecond regime. Representative intensity-dependent TH spectra for selected excitation wavelengths are shown in Figures 5e-h. When the excitation wavelength lies far from the resonance, increasing the input intensity produces negligible changes in the spectral shape of the TH emission (Figures 5e and 5h). Conversely, when the pump wavelength is tuned close to resonance, increasing the intensity leads to pronounced spectral reshaping and broadening of the TH signal (Figures 5f and 5g). Note that the corresponding femtosecond TH efficiency curves at these selected wavelengths, as a function of input intensity, are reported in Sec. S7 of the Supplementary Information.

These observations indicate that, under femtosecond excitation, nonlinear self-action manifests primarily through intensity-dependent modifications of the spectral properties of the generated TH emission.

## DISCUSSION AND CONCLUSION

In summary, we have shown that nonlinear self-action in resonant dielectric metasurfaces manifests in two complementary ways, depending on the temporal characteristics of the excitation. Under spectrally narrow picosecond pumping, the qBIC resonance is driven coherently over its characteristic lifetime, yielding strong enhancement of the TH conversion efficiency together with pronounced deviations from the cubic power scaling expected for a third-order process.

In contrast, when the metasurface is excited by broadband femtosecond pulses, the interaction enters a transient regime in which a continuum of spectral components couples to the resonance with different efficiencies. In this case, the nonlinear response is encoded in the spectral structure of the emitted harmonic: intensity-dependent spectral reshaping and broadening of the TH spectrum arise from the dynamic filtering imposed by the ultrafast, intensity-dependent evolution of the qBIC resonance.

A central outcome of this work is that the observed spectral modulation and deviations from cubic scaling can be consistently explained by an intensity-dependent modification of the resonant response induced by the strong electromagnetic confinement within the qBIC mode. This interpretation is supported by the excellent agreement between experiment and simulations. Additional simulations assessing the role of two-photon absorption show that the spectral shift is predominantly driven by the Kerr-induced refractive-index change, while two-photon absorption has only a minor influence on the resonance position.

This intensity-dependent modification of the resonant response manifests differently in the two temporal excitation regimes. Under narrowband picosecond excitation, resonant field confinement enhances the contribution of higher-order nonlinear susceptibilities to the optical response. At sufficiently high local field intensities, fifth- and seventh-order nonlinearities become increasingly relevant in addition to the conventional third-order response, contributing to the nonlinear polarization at both the fundamental and third-harmonic frequencies. Because these susceptibilities are complex, their combined action modifies both the amplitude and phase of the nonlinear polarization, including dispersive and absorptive contributions, and leads to the pronounced reduction of the scaling exponent of the generated TH signal. This behavior is well reproduced by our CW simulations, which include the nonlinear polarization up to seventh order. Under broadband femtosecond excitation, the response to the same field confinement manifests instead as spectral reshaping of the emitted TH signal. Already at low input intensity, the signature of the resonance is encoded in the spectral features of the TH emission. The transient interaction between the short pulse and the resonance requires a full time-domain treatment to capture the underlying dynamics. Using our hydrodynamic model, we reproduce the spectral features observed in the TH signal, which become more pronounced as the pump spectrum overlaps with the qBIC resonance. Within this framework, the observed intensity-dependent spectral modulations are interpreted as arising from an intensity-dependent refractive-index change that dynamically reshapes the qBIC resonance during excitation.

Although a full time-domain treatment could in principle be applied to both excitation regimes, the picosecond regime, because of the narrowband excitation, is well captured by a CW approach, which more directly highlights the mechanisms governing the TH power scaling. Overall, our results establish a unified framework for nonlinear self-action in resonant metasurfaces, in which strong field confinement simultaneously governs conversion efficiency, power-law scaling under narrowband excitation, and spectral dynamics under broadband excitation.

Although the present study focuses on a qBIC metasurface, the underlying physical picture is not restricted to this specific resonance architecture. Nonlinear self-action results from the intensity-dependent modification of a resonant optical response and is therefore expected to extend to other resonant metasurfaces supporting sufficient field confinement, including guided-mode and quasi-guided resonances. Importantly, the effects reported here are already clearly observable without relying on an extremely high-Q resonance. Systems with stronger field enhancement and narrower linewidths can exhibit even greater sensitivity to nonlinear refractive-index changes. The distinction between quasi-stationary narrowband excitation and transient broadband excitation therefore provides a framework applicable to a broader class of resonant nonlinear metasurfaces.

More broadly, our findings highlight that when ordinary dielectric materials are driven into regimes of strong confinement and ultrafast nonlinear response, they naturally exhibit time-dependent behavior. The resonance reshaping, modified coupling conditions, and intensity-dependent spectral signatures observed here do not require externally engineered modulation or exotic material platforms but arise directly from the intrinsic dynamics of resonant nanostructures.

By clarifying how distinct temporal excitation regimes, through their different bandwidths and coupling conditions, shape the observable signatures of nonlinear self-action, this work provides a consistent framework for interpreting nonlinear dynamics in resonant metasurfaces and offers practical guidance for the design of ultrafast nanophotonic devices. At the same time, it places qBIC-mediated self-action within the broader context of time-dependent photonics, highlighting that many effects typically associated with externally modulated systems can emerge naturally from the intrinsic nonlinear dynamics of strongly driven resonant media.

Looking ahead, the ability to access complementary signatures of nonlinear self-action through both narrowband and broadband excitation opens new opportunities for tailoring ultrafast responses in resonant dielectric metasurfaces. By controlling the temporal and spectral structure of the pump, for example through chirping, multi-color excitation, pump–probe schemes, or pulse shaping, it becomes possible to selectively enhance or suppress different components of the dynamic resonance modulation, thereby shaping the spectral response of the emitted harmonics. Moreover, the intrinsic time-dependent behavior discussed here positions qBIC metasurfaces as compact platforms for exploring ultrafast spatio-temporal phenomena and for engineering nonlinear photonic functionalities in inherently time-varying systems.

**Data availability**. The data that support the findings of this study are available from the corresponding author upon reasonable request.

**Acknowledgements**. M.A.V., M.B. and M.C. acknowledge financial support from the Italian Ministry of University and Research (MUR) through PRIN grants No. 2022YJ5AZH and 2022BC5BW5. A.N. and M.C. acknowledge financial support from the European Union's Horizon Europe research and innovation programme under the Marie Sklodowska-Curie Actions HORIZON-MSCA-2024-PF-01-01 (grant agreement No. 101203711). A.N. was partially funded by the project NQSTI—ID PE_00000023 funded by the European Union under the NextGenerationEU program - CUP H43C22000870001 Spoke 6.

**Author contributions**. M.A.V. and M.B. conceived the project. S.F., J.O., S.M. and M.B. fabricated and characterized the samples. A.N., A.D.F., A.Z., M.F. and M.C. designed and performed the experiments. M.S. and M.A.V. carried out numerical simulations. A.N. and A.D.F. analyzed the data. A.N., M.F., M.C., M.S. and M.A.V. contributed to the theoretical analysis and interpretation of the results. M.F., M.C., M.B. and M.A.V. supervised the research. A.N. drafted the manuscript with contributions from M.S., M.F., M.C. and M.A.V. All authors discussed the results and contributed to the final version of the manuscript.

**Conflict of Interest**. The authors declare no conflict of interest.

**Supplementary information**.
S1: Linear Transmission Setup
S2: TE and TM Polarization Comparison
S3: Picosecond Experimental Setup and Power Stability
S4: Picosecond Power-Law Fits
S5: Picosecond Efficiency Curves
S6: Femtosecond Experimental Setup
S7: Femtosecond Efficiency Curves
S8: Nonlinear Model for Third-Harmonic Generation
S9: Role of Kerr Nonlinearity and Two-Photon Absorption in the Spectral Shift
Figures S1-S9

## MATERIALS AND METHODS

**Sample fabrication.** The metasurfaces are fabricated on a glass wafer coated with a silicon thin film deposited by chemical vapor deposition (CVD). The final film thickness is 140 nm. Owing to deposition on a dielectric substrate, the silicon film is amorphous. Arrays of rectangular nanobars parallel to the substrate edge are defined by electron-beam lithography (EBL), followed by reactive ion etching (RIE). Each unit cell consists of two bars with different widths, 147 nm and 190 nm, respectively, both having a length of about 150 µm. The gap between the two bars is about 255 nm. The unit cell is periodically repeated over a 200 $\mu$m × 200 $\mu$m area with a lattice period of 770 nm. A negative resist is spin-coated onto the amorphous silicon film and exposed according to the designed pattern using an EBL system Crestec AP50RD with 50 kV acceleration voltage and 50 pA beam current. Specifically, we use an Allresist ARN-7520 ebeam resist covered with a thin layer of Espacer 300Z to avoid charging. The exposure dose is 40 µC/cm$^2$. After exposure, the anticharging layer is removed by immersing the sample 20 s in $H_2O$ and the resist is developed in Allresist Developer AR 300-47 solution for 1 min followed by 10 s rinse in $H_2O$ to quench the development. The pattern is subsequently transferred into the Si thin film by reactive ion etching. ICP-RIE Sentech SI500 is used with the following process parameters: $C_4F_8$ (17 sccm) + $SF_6$ (11.5 sccm), Pressure 0.5 Pa, ICP Power 200 W, RF Power 20 W, electrode/sample temperature setting -20°C with a resulting Si etching rate of 2.09 nm/s.

**Linear optical characterization.** We measure the linear transmission spectra using broadband illumination from a supercontinuum source. A broadband linear polarizer controls the polarization state of the incident light, allowing us to compare excitation with the electric field parallel and orthogonal to the nanobars. The beam is weakly focused onto the metasurface using a long-focal-length lens with focal length $f = 60$ mm. The transmitted light is collected with an objective and sent to an infrared spectrometer, which records the transmission spectrum over the wavelength range of interest. The full setup is shown in Sec. S1 of the Supplementary Information, while the TE/TM comparison is reported in Sec. S2.

**Picosecond third-harmonic measurements.** Picosecond nonlinear measurements are performed using a tunable laser operated over the 1505–1537 nm wavelength range. The pulse duration is 7 ps, the spectral linewidth is approximately 0.5 nm, and the repetition rate is 50 MHz. A linear polarizer selects the incident polarization. A wedge picks off a small fraction of the beam and sends it to a calibrated photodetector, which monitors the optical power reaching the sample during the wavelength and power sweeps. We use this signal as feedback to adjust the laser output power and stabilize the power at the sample across the full excitation-wavelength range. The remaining beam is weakly focused onto the metasurface with a lens of focal length $f = 60$ mm, and the transmitted light is collected with an objective. The residual fundamental beam is rejected using short-pass spectral filters, and the generated TH signal is detected with a single-photon avalanche diode. The picosecond setup and the corresponding power-stability measurement are reported in Sec. S3 of the Supplementary Information. We also conducted an analysis of the picosecond TH conversion efficiency, estimated from the detected photon count rate after correcting for the collection-path transmission and detector quantum efficiency, as reported in Sec. S5.

**Femtosecond third-harmonic measurements.** Femtosecond nonlinear measurements are performed with a tunable femtosecond laser delivering pulses of approximately 60 fs duration and a spectral bandwidth of about 60 nm. The repetition rate is 1 MHz. The input beam is sent through a linear polarizer and weakly focused onto the metasurface using a long-focal-length lens with $f = 60$ mm. The transmitted light is collected with an objective, spectrally filtered to isolate the TH signal, and coupled into a multimode fiber connected to a visible spectrometer. This allows us to measure the TH spectrum as a function of both excitation wavelength and input intensity. The femtosecond setup is reported in Sec. S6 of the Supplementary Information.

We also conducted an analysis of the femtosecond TH conversion efficiency, estimated from the measured TH spectra after correcting for the wavelength-dependent detection response, as reported in Sec. S7.

**Numerical simulations.** Two complementary numerical approaches are employed in this work, reflecting the different temporal nature of the excitation regimes investigated experimentally.

For the picosecond regime, the narrow excitation linewidth makes the interaction effectively quasi-stationary, so that the nonlinear response can be modeled in the frequency domain using a two-frequency basis containing only the fundamental field at $\omega$ and the generated third-harmonic field at $3\omega$. Within this framework, the nonlinear polarization of amorphous silicon is expanded up to seventh order, including the complex nonlinear susceptibilities $\chi^{(3)}$, $\chi^{(5)}$, and $\chi^{(7)}$. Although higher-frequency components such as $5\omega$ and $7\omega$ are not propagated, the fifth- and seventh-order nonlinearities contribute directly to the nonlinear polarization at both $\omega$ and $3\omega$. The pump-only contribution driving third-harmonic generation can therefore be written as

$$P_{\mathrm{THG}} = \varepsilon_0 E_\omega^3 \left[\chi^{(3)} + 5\chi^{(5)}|E_\omega|^2 + 21\chi^{(7)}|E_\omega|^4\right], \tag{1}$$

where $P_{\mathrm{THG}}$ is the nonlinear polarization component driving third-harmonic generation at $3\omega$, $\varepsilon_0$is the vacuum permittivity, $E_\omega$ is the complex amplitude of the electric field at the fundamental frequency $\omega$, and $\chi^{(3)}$, $\chi^{(5)}$, and $\chi^{(7)}$ are the third-, fifth-, and seventh-order nonlinear susceptibilities of amorphous silicon, respectively. Because the susceptibilities are complex, the higher-order terms modify both the amplitude and phase of the THG source and introduce intensity-dependent dispersive and absorptive contributions, thereby accounting for the observed deviation from cubic scaling and the saturation of the TH response. Within the same CW framework, we performed additional simulations with and without the two-photon absorption contribution, in order to assess the origin of the intensity-dependent spectral shift. The full expressions for the nonlinear polarization at the fundamental and third-harmonic frequencies are reported in Sec. S8 of the Supplementary Information.

For femtosecond excitation, instead, the broadband pulse interacts transiently with the resonance, and the emitted TH spectrum evolves dynamically during the pulse. Describing this regime requires a full time-domain treatment, which we implement using a hydrodynamic Maxwell-Lorentz model, formulated in Gaussian units. Within this approach, the macroscopic electromagnetic fields obey Maxwell's equations, while the material response is described by driven, damped oscillators that account for both linear and nonlinear polarization dynamics. For the time-domain femtosecond model, we focus on centrosymmetric media and on bulk third-order nonlinearities; second-order surface and magnetic sources are retained at the level of the equations of motion for completeness, but will not be discussed further, as their contribution is predicted to be too weak to be observed under our conditions.

The dominant contribution to the polarization in amorphous silicon arises from bound electrons. We model their dynamics using a nonlinear Lorentz oscillator driven by the local electric and magnetic fields. The polarization $\boldsymbol{P}(\boldsymbol{r}, t)$ obeys:

$$\frac{\partial^2 \boldsymbol{P}}{\partial t^2} + \gamma \frac{\partial \boldsymbol{P}}{\partial t} + \omega_0^2 \boldsymbol{P} + \alpha(\boldsymbol{P}\cdot\boldsymbol{P})\boldsymbol{P} = \frac{n_b e^2}{m_b}\boldsymbol{E} + \frac{e}{m_b}(\boldsymbol{P}\cdot\nabla)\boldsymbol{E} + \frac{e}{m_b c}\frac{\partial \boldsymbol{P}}{\partial t}\times \boldsymbol{B}. \tag{2}$$

Here $n_b$ is the bound-electron density, $m_b$ the effective mass, $\gamma$ is the damping coefficient, and $\omega_0$ the resonance frequency of the dominant Lorentz oscillator. The coefficient $\alpha$ controls the strength of the third-order nonlinear restoring force and is directly related to the material's third-order susceptibility $\chi^{(3)}$. The right-hand side of Eq. (2) contains three distinct driving contributions: i) the linear dipole term $(n_b e^2/m_b)\boldsymbol{E}$, which reproduces the usual Lorentz model in the absence of nonlinearities; ii) the nonuniform-field (or pseudo-convective) term $(e/m_b)(\boldsymbol{P}\cdot\nabla)\boldsymbol{E}$, which arises from spatial variations of the electric field acting on a bound polarization distribution and effectively captures higher-order multipolar corrections without implying true charge convection; iii) the magnetic Lorentz term $(e/m_b c)\ \partial_t \boldsymbol{P}\times\boldsymbol{B}$, which accounts for the coupling between the bound current density $\partial_t \boldsymbol{P}$ and the magnetic field. In centrosymmetric media, the combination of the nonlinear restoring force $\alpha(\boldsymbol{P}\cdot\boldsymbol{P})\boldsymbol{P}$ and the linear dipole term generates the bulk third-order nonlinear polarization responsible for TH generation. The Coulomb and magnetic Lorentz terms can, in principle, contribute to both second- and third-order processes, but in the present work we do not analyze their individual contributions in detail, as the corresponding signals are expected to be below the detection threshold.

In the weak-field limit, neglecting the nonlinear term $\alpha(\boldsymbol{P}\cdot\boldsymbol{P})\boldsymbol{P}$ and the Coulomb and magnetic contributions, Eq. (2) reduces to:

$$\frac{\partial^2 \boldsymbol{P}}{\partial t^2}+\gamma\frac{\partial \boldsymbol{P}}{\partial t}+\omega_0^2\boldsymbol{P}=\frac{n_b e^2}{m_b}\boldsymbol{E}. \quad (3)$$

which yields the familiar linear susceptibility $\chi^{(1)}(\omega)$ and dielectric function $\epsilon(\omega)=1+4\pi\chi^{(1)}(\omega)$. Retaining the nonlinear restoring force and performing a perturbative expansion in the field amplitude leads to an effective third-order susceptibility $\chi^{(3)}(3\omega;\omega,\omega,\omega)$ that inherits the dispersive properties of the underlying Lorentz oscillator. In the time-domain formulation used here, we do not explicitly separate homogeneous and inhomogeneous solutions; instead, the full nonlinear polarization dynamics, including TH generation, emerges directly from the integration of the coupled system formed by Eq. (2) and Maxwell's equations $\nabla\times\mathbf{E}=-\frac{\mathbf{1}}{c}\frac{\partial\mathbf{B}}{\partial t}$ and $\nabla\times\mathbf{B}=\frac{\mathbf{1}}{c}\frac{\partial\mathbf{E}}{\partial t}+\frac{4\pi}{c}\frac{\partial\mathbf{P}}{\partial t}+\frac{4\pi}{c}\mathbf{J}_{\text{free}}$. In the time domain, this operation is performed using a split-step scheme. Free-space propagation is handled by advancing the fields according to Maxwell's equations with $\boldsymbol{P}$ held fixed, typically using Fourier methods in space. The material response is then updated by integrating Eq. (2) at each spatial point, using a second-order accurate predictor–corrector method. This procedure preserves both linear and nonlinear dispersion and remains numerically stable in the presence of strong field localization and narrow resonances associated with qBIC modes. In this way, the hydrodynamic Maxwell–Lorentz model provides a compact yet comprehensive framework to describe TH generation in amorphous silicon structures, while keeping explicit track of the Coulomb and magnetic Lorentz contributions at the level of the equations of motion.

# Supplementary Information:

# Nonlinear Self-Action across Temporal Regimes in Resonant Dielectric Metasurfaces

Alfonso Nardi,[1] Sonia Freddi,[2] Michael Scalora,[3] Agostino Di Francescantonio,[1] Johann Osmond,[4] Sofia Martins,[4] Attilio Zilli,[1] Marco Finazzi,[1] Michele Celebrano,[1] Monica Bollani,[2] Maria Antonietta Vincenti[5]

[1]Department of Physics, Politecnico di Milano, Piazza Leonardo 32, 20133 Milano, Italy
[2]Institute of Photonic and Nanotechnology (IFN) - Consiglio Nazionale delle Ricerche (CNR), Piazza Leonardo 32, 20133 Milano, Italy
[3]Institute of Photonics and Quantum Sciences, Heriot-Watt University, SUPA, Edinburgh EH144AS, United Kingdom
[4]ICFO - Institut de Ciències Fotòniques, the Barcelona Institute of Science and Technology, Castelldefels Barcelona 08860, Spain
[5]Department of Information Engineering, Università degli Studi di Brescia, Via Branze 38, 25123 Brescia, Italy

This document provides supplementary information for the article "Nonlinear Self-Action across Temporal Regimes in Resonant Dielectric Metasurfaces". Section S1 shows the experimental setup used for linear transmission measurements, while Sec. S2 reports the comparison between transverse-electric (TE) and transverse-magnetic (TM) excitation. In Sec. S3, we report the picosecond experimental setup and the corresponding power-stability measurement during the wavelength sweep. Section S4 reports the power-law fits used to characterize the nonlinear response as a function of input intensity under picosecond excitation. Section S5 presents the third-harmonic (TH) conversion efficiency curves measured under picosecond excitation. Sections S6 and S7 show the femtosecond experimental setup and the corresponding TH conversion efficiency curves. Section S8 describes the nonlinear model used to calculate third-harmonic generation under picosecond excitation, while Sec. S9 assesses the relative role of Kerr nonlinearity and two-photon absorption in the simulated intensity-dependent spectral shift.

## S1 Linear Transmission Setup

The linear-transmission measurement procedure is described in the Materials and methods section of the main text. In Fig. S1, we report the corresponding experimental setup.

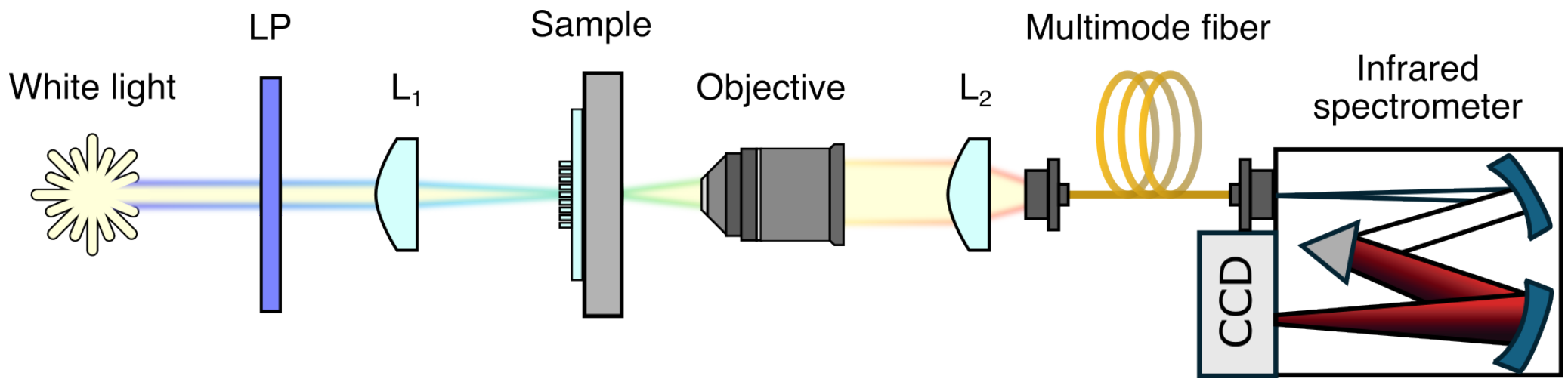

Figure S1: Experimental setup used for linear transmission spectral measurements under white light illumination. White light: supercontinuum laser source; LP: Linear Polarizer; L: Lens.

## S2 TE and TM Polarization Comparison

To assess the polarization dependence of the resonant response of our metasurface, we compare the linear transmittance and TH signal measured under transverse-electric (TE) and transverse-magnetic (TM) excitation, where the TE mode corresponds to the polarization direction along the amorphous silicon nanobars, while the TM mode corresponds to the polarization direction perpendicular to the nanobars. Figure S2 reports the result of the linear characterization, shown on the left axis, together with the nonlinear characterization under picosecond excitation, shown on the right axis. By comparing the linear transmittance spectra, we observe that TE excitation gives rise to the Fano resonance shape associated with the qBIC mode (Fig. S2a), while TM excitation leads to a flat transmittance over the entire wavelength range (Fig. S2b). Under TE excitation, the TH signal is strongly enhanced at the Fano dip. By contrast, under TM excitation, we do not observe any measurable TH signal.

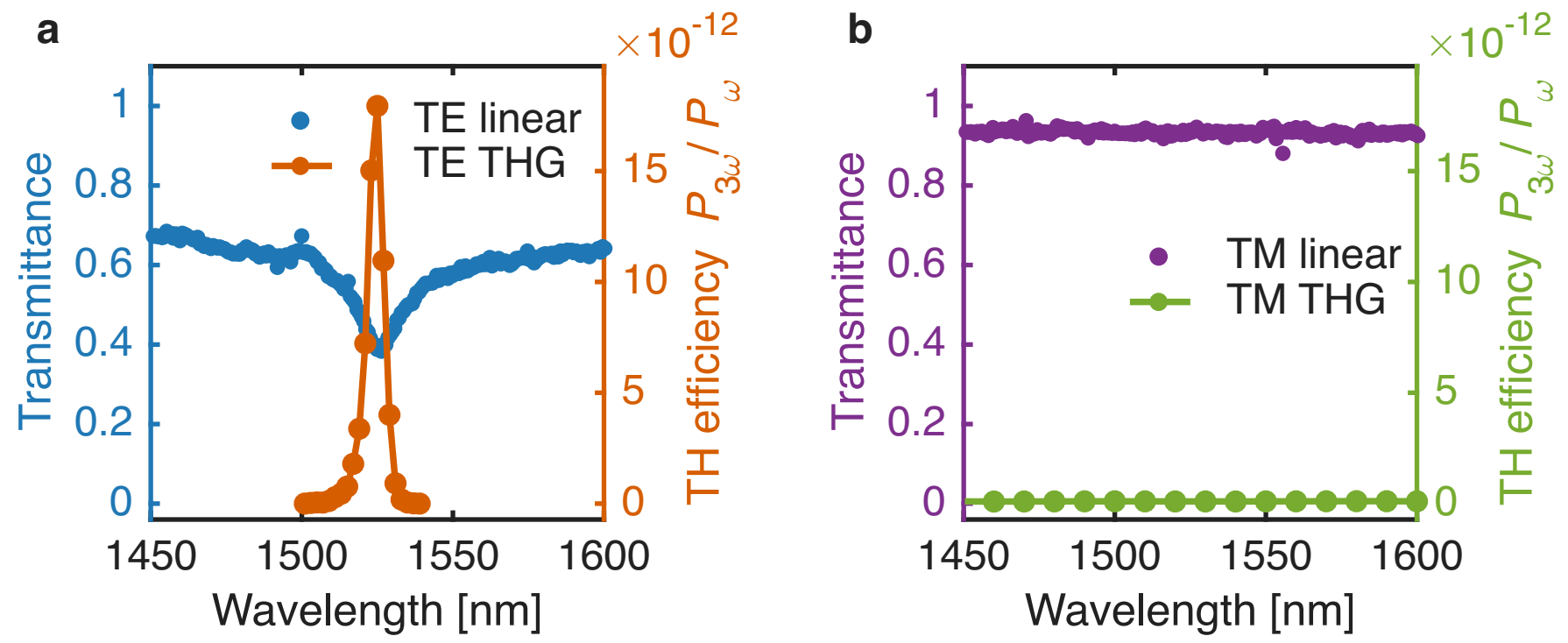

Figure S2: Polarization-dependent linear and nonlinear optical response of the metasurface. Linear transmittance and estimated TH conversion efficiency are compared for (a) TE and (b) TM excitation. The TH efficiency is reported as $P_{3\omega}/P_\omega$ after correcting the detected counts for the wavelength-dependent collection-path transmission and detector response (see Sec. S5).

## S3 Picosecond Experimental Setup and Power Stability

The picosecond nonlinear measurement procedure is described in the Materials and methods section of the main text. Here, we report the corresponding experimental setup and the power-stability measurement during the wavelength sweep.

### S3.1 Picosecond setup

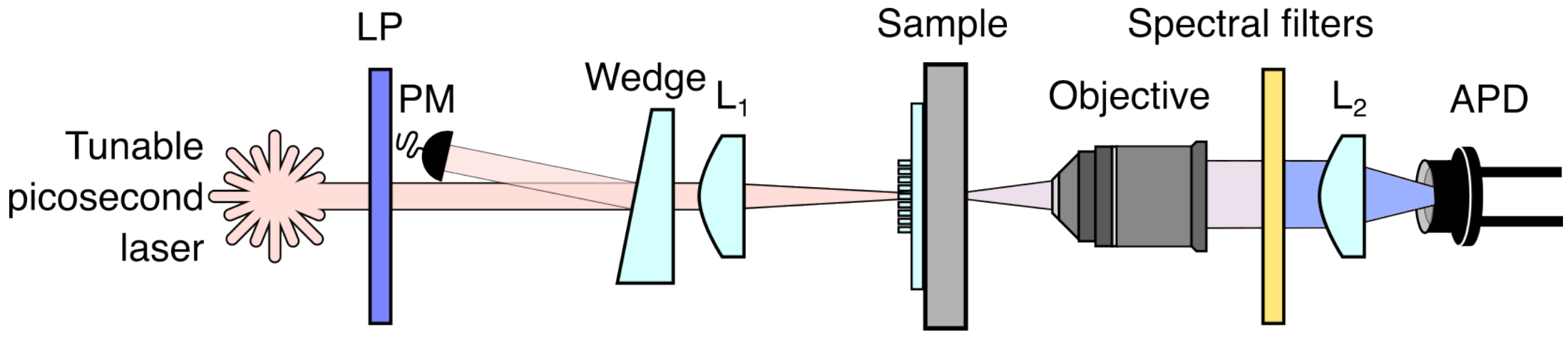

Figure S3: Experimental setup used for nonlinear measurements under picosecond pulse illumination. LP: Linear Polarizer; L: Lens; APD: Avalanche Photodiode.

### S3.2 Power monitoring and stability during the wavelength sweep

During the wavelength sweep, a fraction of the incident beam is monitored with a calibrated photodetector and used as feedback to stabilize the power at the sample. Figure S4 reports the measured power at the sample and the corresponding normalized TH response over the excitation-wavelength range used in the picosecond measurements.

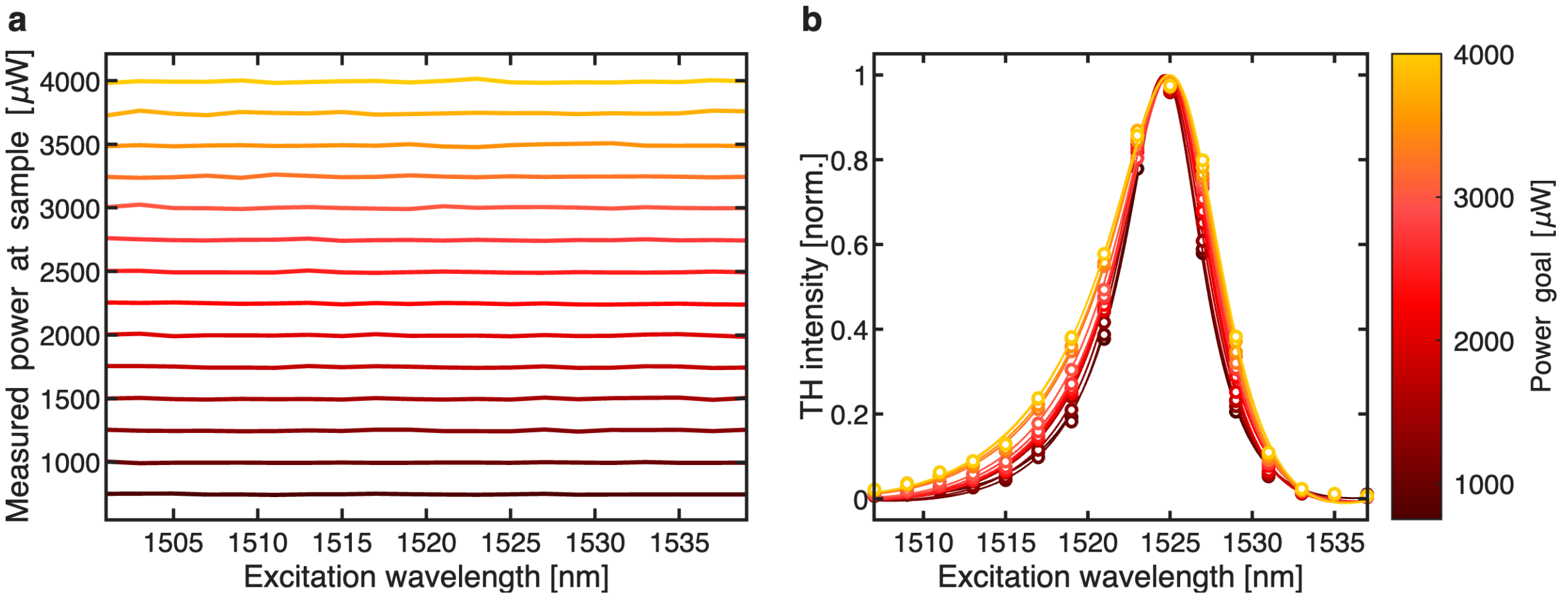

Figure S4: Picosecond power stability and wavelength-dependent TH response. (a) Measured power at the sample as a function of excitation wavelength for the picosecond power sweep. (b) Normalized TH intensity spectra measured over the same excitation-wavelength range for all input powers. The color scale indicates the nominal power goal used for each sweep.

## S4 Picosecond Power-Law Fits

This section describes the procedure used to extract the power-law exponent of the picosecond third-harmonic response from the measured data. We perform the analysis independently at each excitation wavelength, allowing us to obtain the wavelength-dependent exponent $\alpha$, which characterizes the scaling of the TH intensity with input intensity. The procedure is as follows: for each wavelength, the TH signal is normalized to its value at the lowest input power and fitted on a log-log scale with a power law:

$$I_{\mathrm{TH}} \propto I_{\mathrm{in}}^{\alpha} .$$

The fit is performed in two different windows: a low-intensity window, where we fix the exponent to the cubic value $\alpha = 3$, which serves as a reference for the expected low-intensity cubic scaling; and a high-intensity window, where we let the exponent be a free parameter. This allows us to clearly visualize the deviation from cubic scaling, which becomes stronger as the excitation wavelength approaches the maximum of the TH conversion efficiency. This procedure therefore provides the slope extracted from the high-intensity fit, which is the quantity used in Fig. 4d of the main text.

Figure S5 shows the individual fits used to extract the slope $\alpha$. The empty circular markers represent the data points used for the reference cubic fit at low intensity, while the filled square markers highlight the data points shown in the high-intensity fitting range. The solid black line represents the high-intensity fit, while the dashed gray line represents the low-intensity cubic fit used as a reference. The exponent $\alpha$ extracted from the high-intensity fit is indicated in each panel.

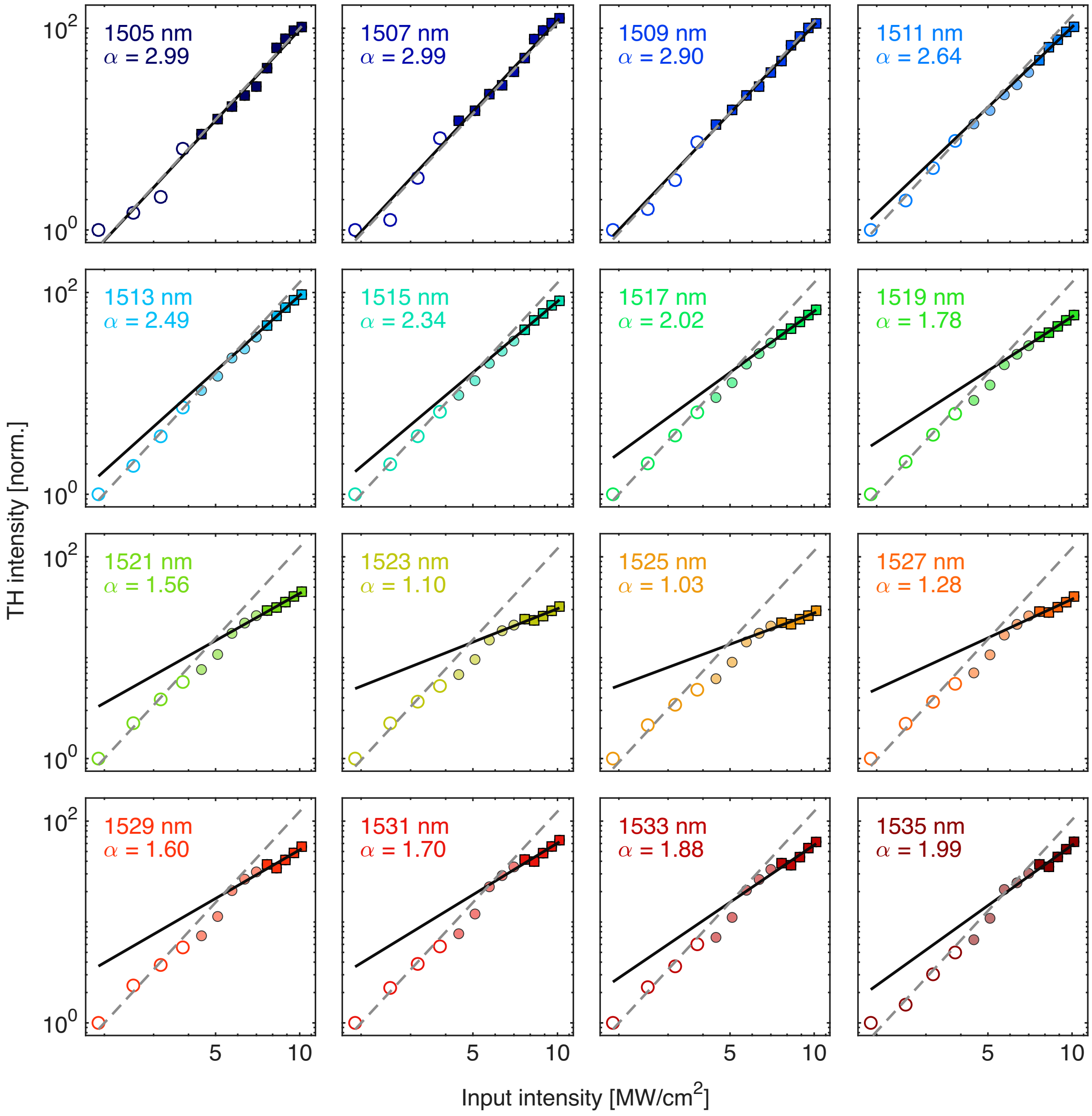

Figure S5: Individual picosecond power-law fits for each excitation wavelength. Each panel shows the normalized TH intensity as a function of input intensity on a log-log scale. The solid black line is the high-intensity fit, while the dashed gray line is the low-intensity cubic fit used as a reference. The exponent $\alpha$ extracted from the high-intensity fit is indicated in each panel. Empty circular markers represent the data points used for the low-intensity cubic fit, while filled square markers indicate the data points considered for the high-intensity fit.

# S5 Picosecond Efficiency Curves

In this section, we report the picosecond TH efficiency for the full dataset, comprising different excitation wavelengths and input intensity values. The TH efficiency is estimated as follows. First, the detected TH count rate is corrected for the collection-path transmission. In the wavelength range considered here, the generated TH wavelength lies between 501.7 nm and 512.3 nm. In this range, we characterized the objective transmission to be approximately 88%, while the transmission of each Thorlabs FGB39 spectral filter is about 86%. Since three FGB39 filters are used, the total filter transmission is about 64%. We then correct the detected TH count rate for the SPAD quantum efficiency at the TH wavelength, which is 46–48% over the same spectral range [1]. The resulting corrected photon rate is then converted into optical power by multiplying it by the photon energy at the TH wavelength. Finally, the estimated TH power is divided by the measured average input power at the sample to obtain the TH conversion efficiency ($P_{3\omega}/P_{\omega}$). We report the resulting TH efficiency as a function of input intensity for all excitation wavelengths in Fig. S6.

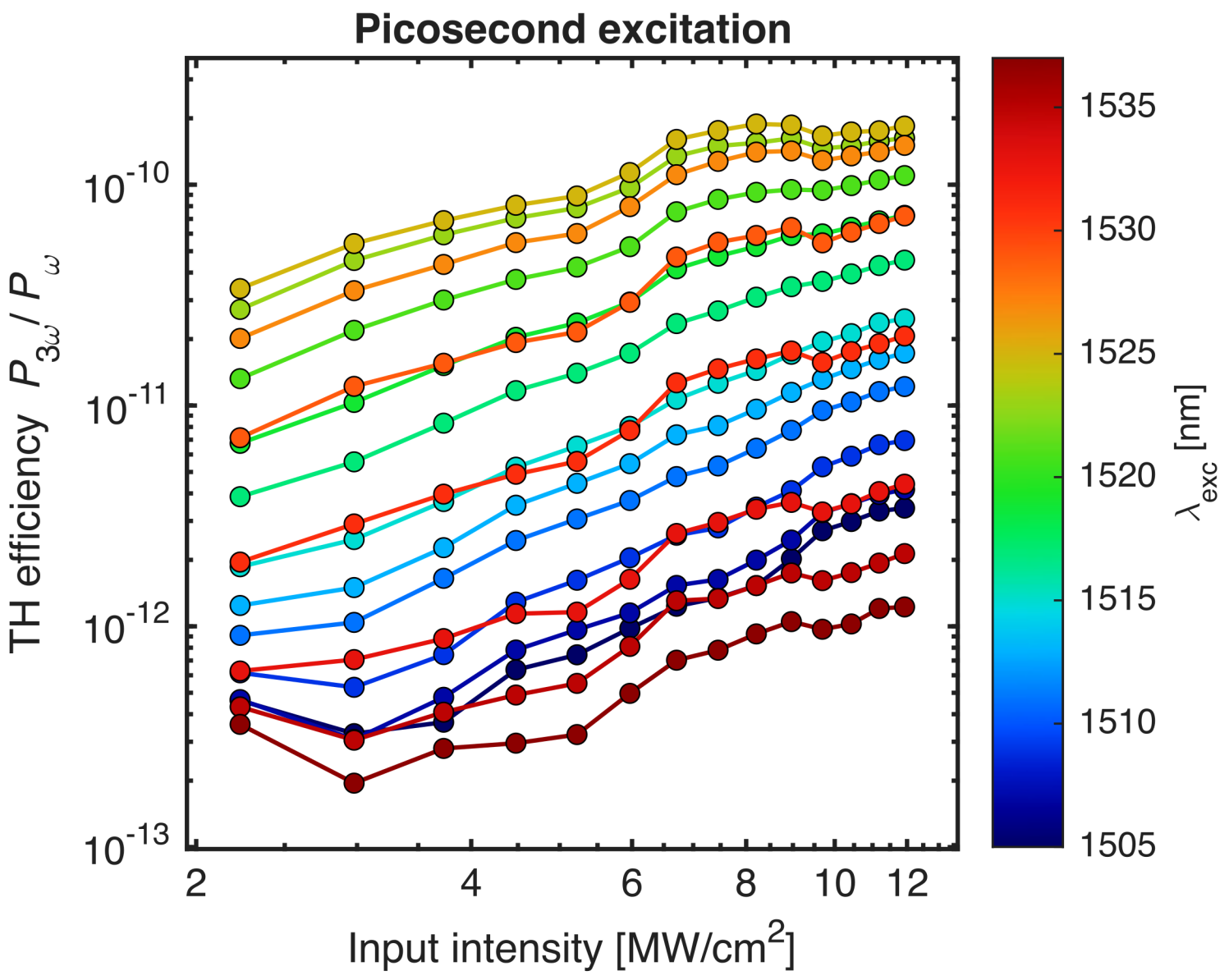

Figure S6: Estimated picosecond TH conversion efficiency as a function of input peak intensity for all excitation wavelengths from 1505 nm to 1537 nm. The efficiency is calculated as $P_{3\omega}/P_{\omega}$ after correcting the detected TH counts for the collection-path transmission and detector efficiency at the generated TH wavelength.

# S6 Femtosecond Experimental Setup

The femtosecond nonlinear measurement procedure is described in the Materials and methods section of the main text. In Fig. S7, we report the corresponding experimental setup.

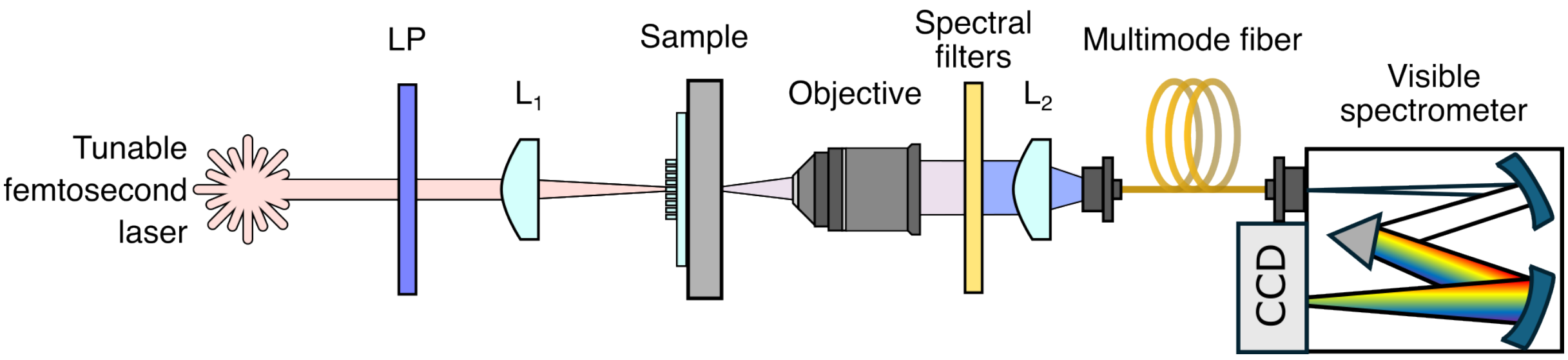

Figure S7: Experimental setup used for femtosecond nonlinear measurements. A tunable femtosecond source is weakly focused onto the metasurface after passing through a linear polarizer. The collected light is spectrally filtered, and a visible spectrometer measures the TH spectrum. LP: Linear Polarizer; L: Lens.

# S7 Femtosecond Efficiency Curves

In this section, we estimate the TH conversion efficiency under femtosecond pulse excitation. We characterize the TH conversion efficiency as a function of input intensity for the four central excitation wavelengths highlighted in the main text: 1480 nm, 1510 nm, 1540 nm, and 1570 nm. We estimate the conversion efficiency as follows. First, we correct the measured TH spectra for the wavelength-dependent CCD response, grating efficiency, objective transmission, exposure time, and neutral-density filter transmission, and then integrate the detected TH signal over the corresponding spectral window. Over the TH spectral windows used in the analysis, the CCD response is about 90–96%, the grating efficiency is about 73%, and the objective transmission is about 87–89%. The neutral-density transmissions and the exposure time depend on the input power, and are chosen in such a way as to ensure similar count rates across the different input intensity values, while avoiding saturation of the CCD and keeping a similar signal-to-noise ratio across the dataset. Finally, the estimated TH power is divided by the measured average input power at the sample to obtain the TH conversion efficiency, defined as $P_{3\omega}/P_{\omega}$.

We show the resulting efficiency curves in Fig. S8, as a function of input intensity for the four excitation wavelengths.

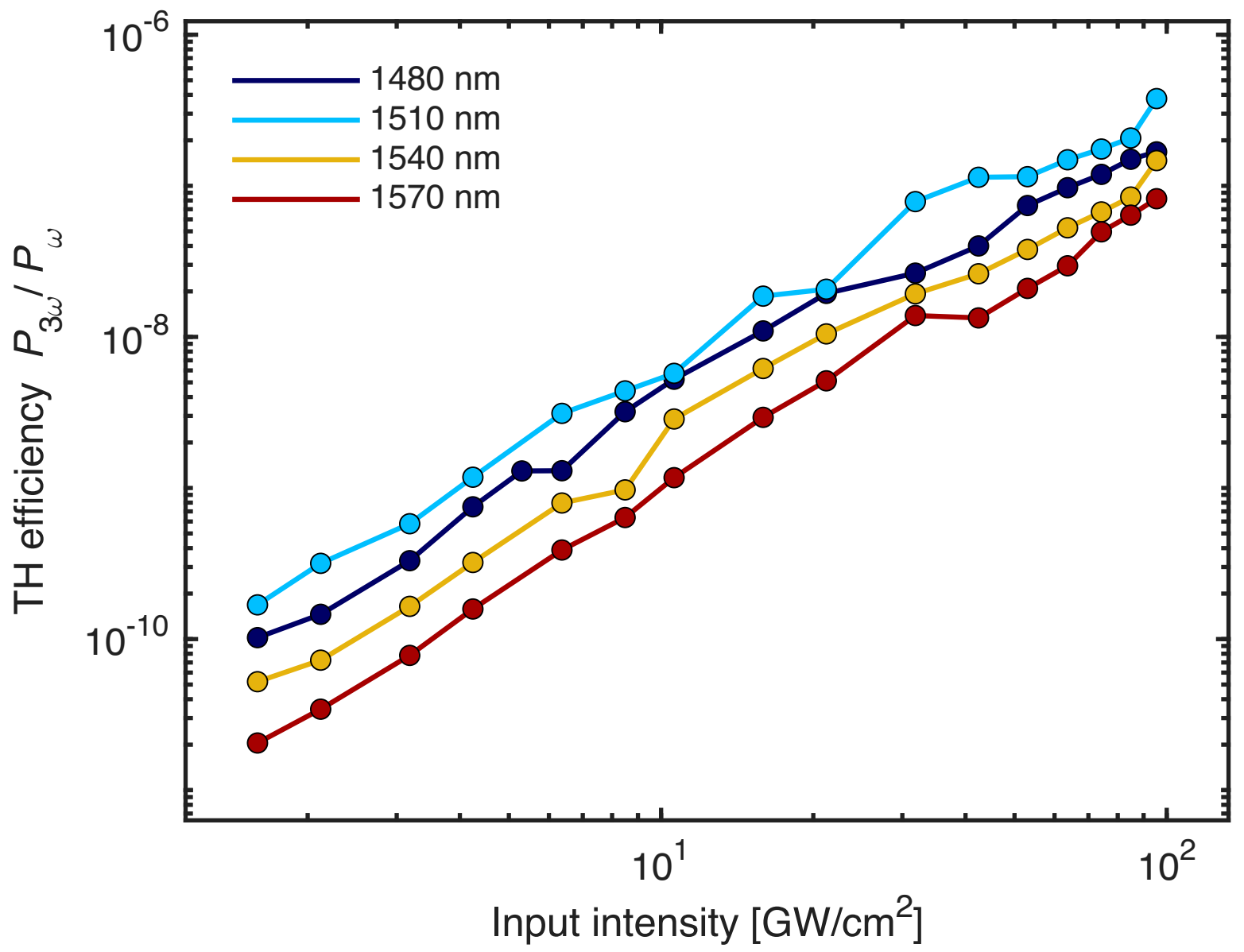

Figure S8: Estimated femtosecond TH conversion efficiency as a function of input intensity for four different excitation wavelengths across the available tunability range (1480 nm, 1510 nm, 1540 nm, and 1570 nm). We calculate the efficiency as $P_{3\omega}/P_{\omega}$ after correcting the detected TH spectra for the wavelength-dependent CCD response, grating efficiency, objective transmission, exposure time, and neutral-density filter transmission.

## S8 Nonlinear Model for Third-Harmonic Generation

This section provides the detailed formulation of the nonlinear frequency-domain model used to describe third-harmonic generation under picosecond excitation.

We consider a TE-polarized electromagnetic field whose only nonzero electric-field component is oriented along the $z$ direction. The field basis is restricted to the fundamental frequency and the third harmonic:

$$\mathbf{E}(\mathbf{r},t) = \hat{\mathbf{z}} \left[ E_{\omega,z}(\mathbf{r}) e^{-i\omega t} + E_{3\omega,z}(\mathbf{r}) e^{-i3\omega t} + \text{c.c.} \right]. \tag{S1}$$

For compactness, we define

$$E_{\omega} \equiv E_{\omega,z}, \qquad E_{3\omega} \equiv E_{3\omega,z}. \tag{S2}$$

The scalar electric field is therefore

$$E_z(\mathbf{r},t) = E_\omega(\mathbf{r})e^{-i\omega t} + E_{3\omega}(\mathbf{r})e^{-i3\omega t} + \text{c.c.} \tag{S3}$$

Amorphous silicon is centrosymmetric and macroscopically isotropic. The nonlinear polarization is truncated at seventh order and is written as

$$\mathbf{P}_{\text{NL}}(\mathbf{r},t) = \hat{\mathbf{z}}P_{\text{NL},z}(\mathbf{r},t), \tag{S4}$$

with

$$P_{\text{NL},z}(\mathbf{r},t) = \varepsilon_0\left[\chi^{(3)}E_z^3(\mathbf{r},t) + \chi^{(5)}E_z^5(\mathbf{r},t) + \chi^{(7)}E_z^7(\mathbf{r},t)\right]. \tag{S5}$$

Because only the $z$-polarized component is present, the relevant tensor contractions reduce to scalar susceptibility coefficients. All susceptibilities entering the model are complex and dispersive. The nonlinear polarization is decomposed only into components oscillating at $\omega$ and $3\omega$:

$$P_{\text{NL},z}(\mathbf{r},t) = P_{\text{NL},\omega,z}(\mathbf{r})e^{-i\omega t} + P_{\text{NL},3\omega,z}(\mathbf{r})e^{-i3\omega t} + \text{c.c.} \tag{S6}$$

Substitution of Eq. (S3) into Eq. (S5) generates all frequency-mixing terms associated with the two retained field components. Since the third-harmonic amplitude remains smaller than the pump amplitude, we retain all pump-only contributions and all terms that are linear in $E_{3\omega}$ or $E_{3\omega}^*$. Terms containing two or more factors of the third-harmonic field are neglected.

## S8.1 Polarization at the fundamental frequency

Within this approximation, the nonlinear polarization at the pump frequency is

$$\begin{aligned} P_{\text{NL},\omega,z} = \varepsilon_0\Big\{&\chi^{(3)}\left[3|E_\omega|^2E_\omega + 3(E_\omega^*)^2E_{3\omega}\right] \\ &+ \chi^{(5)}\left[10|E_\omega|^4E_\omega + 20|E_\omega|^2(E_\omega^*)^2E_{3\omega} + 5E_\omega^4E_{3\omega}^*\right] \\ &+ \chi^{(7)}\left[35|E_\omega|^6E_\omega + 105|E_\omega|^4(E_\omega^*)^2E_{3\omega} + 42|E_\omega|^2E_\omega^4E_{3\omega}^*\right]\Big\}. \end{aligned} \tag{S7}$$

Since $\chi^{(3)}$, $\chi^{(5)}$, and $\chi^{(7)}$ are complex, Eq. (S7) includes both the intensity-dependent phase shift and the intensity-dependent absorption experienced by the pump. We note that the terms in Eq. (S7) couple the third-harmonic field back to the pump and account for pump depletion, back-conversion, and coherent energy exchange between the two frequencies.

## S8.2 Polarization at the third-harmonic frequency

The nonlinear polarization at the third-harmonic frequency is

$$\begin{aligned} P_{\text{NL},3\omega,z} = \varepsilon_0\Big\{&\chi^{(3)}\left[E_\omega^3 + 6|E_\omega|^2E_{3\omega}\right] \\ &+ \chi^{(5)}\left[5|E_\omega|^2E_\omega^3 + 30|E_\omega|^4E_{3\omega}\right] \\ &+ \chi^{(7)}\left[21|E_\omega|^4E_\omega^3 + 140|E_\omega|^6E_{3\omega} + 7E_\omega^6E_{3\omega}^*\right]\Big\}, \end{aligned} \tag{S8}$$

where the pump-only source terms responsible for third-harmonic generation are

$$P_{\text{THG},z} = \varepsilon_0E_\omega^3\left[\chi^{(3)} + 5\chi^{(5)}|E_\omega|^2 + 21\chi^{(7)}|E_\omega|^4\right]. \tag{S9}$$

Because all susceptibility coefficients are complex, the higher-order terms modify both the amplitude and the phase of the THG source. The fifth- and seventh-order contributions also introduce additional intensity-dependent absorption through the imaginary part of the susceptibilities. The saturation of the generated third harmonic therefore results from the combined action of the complex third-, fifth-, and seventh-order contributions and is not determined only by the signs of the real parts of the susceptibilities. It depends on the full complex interference among the nonlinear polarization terms, including both dispersive and absorptive contributions.

We stress that in the numerical implementation, only the fields at $\omega$ and $3\omega$ are propagated. Frequency components such as $5\omega$, $7\omega$, and higher harmonics, which formally arise from the fifth- and seventh-order nonlinearities, are excluded from the field basis. Nevertheless, the contributions of $\chi^{(5)}$ and $\chi^{(7)}$ to the nonlinear polarizations at $\omega$ and $3\omega$ are retained.

The resulting two-frequency model includes the complex nonlinear modification of the pump, nonlinear absorption, pump depletion, back-conversion, and the higher-order saturation of third-harmonic generation.

## S9 Role of Kerr Nonlinearity and Two-Photon Absorption in the Spectral Shift

In this section, we assess the relative role of the dispersive Kerr nonlinearity and two-photon absorption (TPA) in the intensity-dependent spectral shift observed under picosecond excitation. To this end, we perform additional CW simulations to compare the nonlinear response in the presence and absence of TPA. In the model, TPA is represented by the imaginary part of the third-order susceptibility, $\mathrm{Im}\,\chi^{(3)}$. It is therefore suppressed by setting $\mathrm{Im}\,\chi^{(3)} = 0$, while retaining $\mathrm{Re}\,\chi^{(3)}$, which accounts for the dispersive Kerr response. All other nonlinear parameters are kept unchanged between the two calculations.

Figure S9a compares the normalized TH spectra calculated at the minimum and maximum input intensities, $I_{\min} = 1 \times 10^{10}$ W/m$^2$ and $I_{\max} = 1 \times 10^{11}$ W/m$^2$. At both intensity values, the spectra obtained with TPA suppressed ("Kerr") and with TPA included ("Kerr + TPA") nearly overlap. In particular, the intensity-dependent displacement of the spectral maximum is essentially unchanged when TPA is included. This indicates that nonlinear absorption does not appreciably affect the resonance shift, which is mainly governed by the Kerr-induced refractive-index change.

Figure S9b shows the same comparison using the non-normalized TH spectra at the maximum input intensity. This representation highlights the effect of TPA on the magnitude of the generated TH signal, while the spectral position of the resonance remains essentially unchanged. Therefore, within the investigated intensity range, TPA mainly affects the amplitude of the nonlinear response, whereas the intensity-dependent spectral shift is determined primarily by the Kerr contribution.

These results support the interpretation that the observed redshift of the resonantly enhanced TH response originates predominantly from the nonlinear refractive-index change, while absorptive nonlinearities play a secondary role in determining the resonance position.

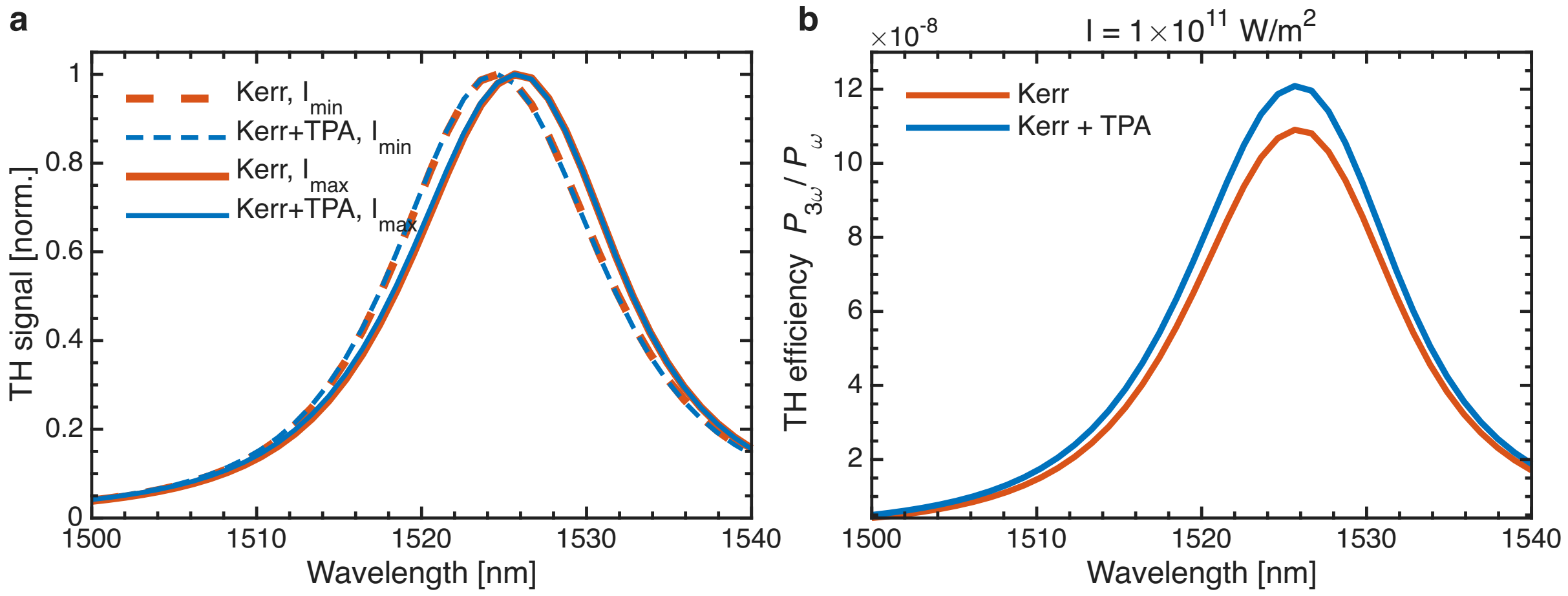

Figure S9: Role of Kerr nonlinearity and two-photon absorption (TPA) in the simulated TH response. (a) Normalized simulated TH spectra calculated with TPA suppressed, $\mathrm{Im}\,\chi^{(3)} = 0$ (Kerr), or using the full complex third-order susceptibility (Kerr + TPA). Dashed lines correspond to $I_{\mathrm{min}} = 1 \times 10^{10}$ W/m$^2$, while solid lines correspond to $I_{\mathrm{max}} = 1 \times 10^{11}$ W/m$^2$. (b) Non-normalized simulated TH spectra at $I_{\mathrm{max}}$, calculated with TPA suppressed (Kerr) or with the full complex $\chi^{(3)}$ (Kerr + TPA).